\documentclass{aa} 
\usepackage{graphicx}
\usepackage{natbib}
\usepackage{txfonts}
\usepackage{amsmath}
\usepackage{graphics,wrapfig,times}
\usepackage[absolute]{textpos}

\begin{document}

\title{Not so typical doubly eclipsing systems}
 \author{Zasche,~P.~\inst{1}}

 \institute{$^{1}$ Charles University, Faculty of Mathematics and Physics, Astronomical Institute, V~Hole\v{s}ovi\v{c}k\'ach 2, CZ-180~00, Praha 8, Czech Republic, \email{petr.zasche@matfyz.cuni.cz} }

 \titlerunning{Not so typical doubly eclipsing systems}
 \authorrunning{Zasche et al.}
 \date{Received \today; accepted ???}

\abstract{Classical doubly eclipsing systems are stars for which two distinct eclipsing periods that apparently come from one point source are detected. Through thorough analysis, the physical parameters of the component stars of these systems can be revealed, and sometimes even the mutual motion of the two doubles around a common barycentre can be constrained. This is most typically done via study of the eclipse-timing variations for both pairs, whose behaviour should be opposite in manner to each other, as the two doubles revolve around a common barycentre. In this study, we present rather unusual systems for which such an easy approach does not work. The period changes are clearly detectable, but only for one pair. These unusual systems are V736 And (periods 0.3596 days and 0.3069 days); ASAS J074939-3037.0 (0.4417 d, 0.2648 d); OGLE BLG-ECL-123507 (1.0170 d, 2.9488 d); ASASSN-V J180818.54-684329.4 (0.3392 d, 3.3234 d); and CRTS J191726.4-543540 (0.3129 d, 2.8178 d).
Several hypotheses are presented regarding the possible origins of such behaviour. However, a final decisive explanation is still yet to be determined.}

 \maketitle

\section{Introduction}

Classical eclipsing binaries are still considered the foundation of modern stellar astrophysics due to their role in deriving highly precise stellar masses, radii, and distances \citep{2012ocpd.conf...51S}. However, during the past decade, a plethora of eclipsing systems with additional eclipses have been found, mainly due to the continuous monitoring of surveys and satellite projects. 

Such systems are either triply eclipsing triples, typically with a rather inflated evolved third component orbiting an inner eclipsing binary on a nearly co-planar orbit (see e.g. \citealt{2022MNRAS.513.4341R} or \citealt{2025A&A...703A.153B}), or  2+2 quadruples with two eclipsing binaries. The difference between these two types of systems is usually clearly visible, even by the naked eye, due to the shape of the eclipses (triply eclipsing triples usually have complicated shapes due to the mutual eclipsing of both inner components). However, the orbital periods of the two doubles in a 2+2 quadruple are often rather long, and detection of its mutual orbit can sometimes be complicated. Consequently, the most common way to prove the gravitational coupling of the two eclipsing pairs is long-term observations of both eclipses of the A and B pairs. Using only this method, we detected several dozen of these mutual orbits in 2+2 doubly eclipsing quadruples and hence proved their quadruple nature (see e.g. \citealt{2023A&A...675A.113Z} or \citealt{2024A&A...687A...6Z}).

In this paper, we present a few systems that differ from those we have described. This includes systems for which the eclipse-timing variation (ETV) diagrams for the A and B pairs do not show opposite behaviours, and hence the mutual A-B orbit cannot be easily derived. As far as I know, no such objects have been reported until now.

\begin{table*}[t]
  \caption{Basic information about the systems.}  \label{systemsInfo}
  \scalebox{0.87}{
  \begin{tabular}{c c c c c c c}\\[-3mm]
\hline \hline\\[-3mm]
  Target name                          &    Other name         &      TESS      &     RA     &     DE      & Mag$_{max}$ & Temperature/sp.type   \\
                                       &                       & identification &  [J2000.0] &  [J2000.0]  &             & information           \\ \hline
 \object{V736 And}                     &   TYC 2798-1339-1     & TIC 432761331  & 00 57 30.9 & +37 38 19.0 &  10.98 (V)  & sp G3$^{\circledast}$      \\ 
 \object{ASAS J074939-3037.0}          &  UCAC4 297-030671     & TIC 150340869  & 07 49 38.9 & -30 36 54.2 &  13.14 (V)  & $T_{eff} = 6281$ K$^{\star}$ \\ 
 \object{OGLE BLG-ECL-123507}          &  UCAC4 304-142251     & TIC 128184860  & 17 49 42.6 & -29 17 05.3 &  14.38 (V)  & $T_{eff} = 7970$ K$^{\star}$ \\ 
 \object{ASASSN-V J180818.54-684329.4} &  TYC 9290-2490-1      & TIC 407731797  & 18 08 18.5 & -68 43 29.3 &  12.09 (V)  & $T_{eff} = 5911$ K$^{\dagger}$ \\ 
 \object{CRTS J191726.4-543540}        & CRTS J191726.4-543540 & TIC 219928818  & 19 17 26.6 & -54 35 39.2 &  13.18 (V)  & $T_{eff} = 5821$ K$^{\#}$ \\  \hline \hline
\end{tabular}}\\
{\small Notes: $^{\circledast}$ - \cite{2023ApJS..264...17Z} ,  $^{\star}$ - \cite{2023A&A...674A...1G} ,  $^{\dagger}$ - \cite{2017AJ....154..259S}, $^{\#}$ - \cite{2025ApJ...984...58H} ,   }
\end{table*}

\section{Data handling and methods used}

For the whole analysis, only the photometric data were used. This is due to the fact that the stars are relatively faint and no usable spectroscopic observations for them were found. Having only photometry available, the analysis is somehow limited in several aspects. The light curve analysis can be done, but the mass ratio can only be roughly estimated or kept fixed at its suggested value. 

However, for the detection of the two binaries' period changes and analysis of their ETV curves, the photometry is good enough. For this purpose, I collected all available past photometry obtained for each star. The changes in a longer period can be determined with a higher degree of conclusiveness with a longer time span. Data from the following surveys or projects were used: 
 \begin{itemize}
     \item ASASsn, the All-Sky Automated Survey for Supernovae \citep{2018MNRAS.477.3145J,2017PASP..129j4502K};
     \item ASAS, the All Sky Automated Survey \citep{1997AcA....47..467P};
     \item WASP, the Wide Angle Search for Planets \citep{2006PASP..118.1407P};
     \item KWS, the Kamogata/Kiso/Kyoto Wide-field Survey \citep{KWS};
     \item ATLAS, the Asteroid Terrestrial-impact Last Alert System \citep{2018AJ....156..241H}; 
     \item Bochum, the Bochum Galactic Disk Survey \citep{2015AN....336..590H}; 
     \item APPLAUSE, the Archives of Photographic PLates for Astronomical USE project of digitalising old photographic plates, available online\footnote{\url{https://www.plate-archive.org/query/}} \citep{2014aspl.conf...53G}; 
     \item KELT, the Kilodegree Extremely Little Telescope \citep{2007PASP..119..923P,2018AJ....155...39O}. 
 \end{itemize}

In addition, new data were used for some of the systems. For several stars, I obtained new observations of one of the pairs to enlarge the time interval covered by the data. These new data are clearly marked in the figures.

The method of analysis of these complicated multiples is similar to our previous studies (see e.g. \citealt{2024A&A...687A...6Z}). For a shorter time interval, I assumed constant linear ephemerides for each binary. Each binary was modelled with the software {\sc{PHOEBE}} \citep{2005ApJ...628..426P}, which is based on the Wilson-Devinney algorithm (see \citealt{1971ApJ...166..605W}). Subtracting the eclipses of one pair, I obtained the photometry only for the other pair, which was analysed in the same way. This process was repeated several times to achieve an acceptable solution (a solution was accepted only when the subtraction of both pairs leads to minimal scatter of the residual points around the zero value). 

Apart from the light curve modelling, the available photometry was also used for deriving the times of eclipses for the particular pair. Only with a large collection of eclipse times for both pairs can one definitively prove the gravitational coupling of both pairs and hence confirm a quadruple 2+2 nature. As the two binaries revolve around a common barycentre, both pairs should show ETVs with an exactly opposite shape (due to the finite speed of light; see e.g. \citealt{1990BAICz..41..231M}). With this method of ETV analysis, we were able to prove the quadruple nature and derive orbital parameters for more than three dozen systems.

 \section{Individual systems}

In the following subsections, I focus on each doubly eclipsing system from our sample. The individual systems were found after careful inspection of all of their available data, but in general they have very little in common. All of them were found during our long-term effort of analysing the published doubly eclipsing systems and collecting (new and existing) data for them for the subsequent ETV analysis. Their identification and some basic information are given in Table \ref{systemsInfo}. As one can see, the stars are dwarf stars, of solar type, or even later-type stars. For all of them, both of their inner orbital periods are shorter than ten days. For most of them, at least one of the binaries is a contact or near-contact type.

\subsection{V736 And}

The very first star is the brightest star in our sample, and it is the only northern hemisphere system. It is named V736~And (it is also known as GSC 02798-01339 and TIC 432761331). It is part of the visual binary WDS J00575+3738AB, its visual components being within $<$~0.5$^{\prime\prime}$. I am not aware of any attempt to derive its visual orbit (only a few observations exist), and no detailed photometric or spectroscopic analysis had been published until now. The only information about its component stars can be inferred from the medium-resolution spectra from the LAMOST (The Large Sky Area Multi-Object Fiber Spectroscopic Telescope) survey, which classifies the system as a G3 spectral type (with a brief remark about its possible stellar activity). 

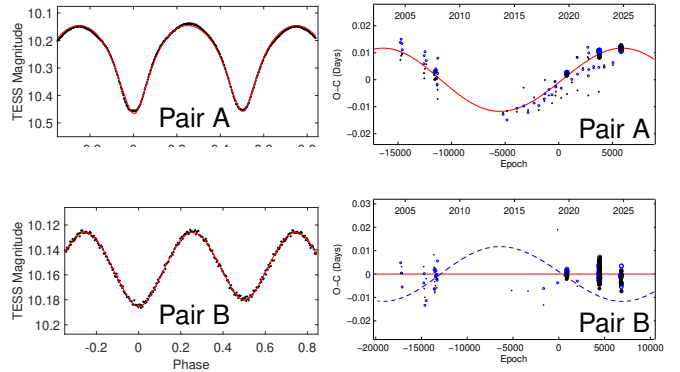
\begin{figure}
  \centering
  \begin{picture}(380,140)
   \put(1,75){\includegraphics[width=0.22\textwidth]{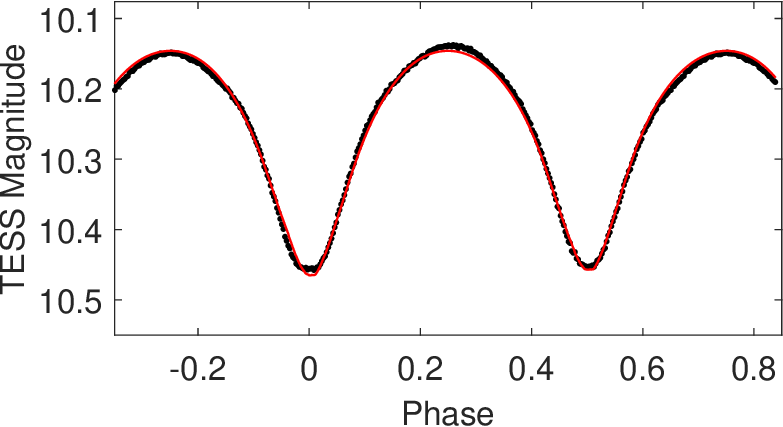}}
   \put(1,1){\includegraphics[width=0.22\textwidth]{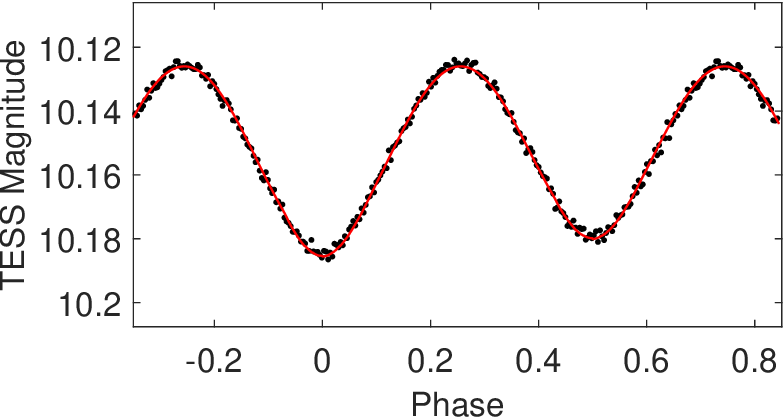}}
   \put(120,1){\includegraphics[width=0.243\textwidth]{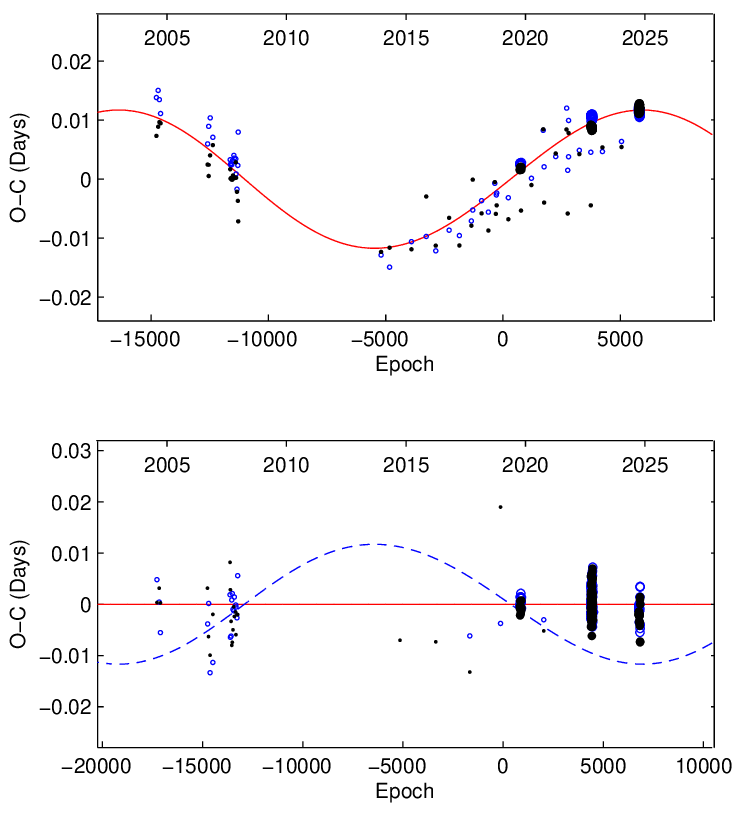}}
    \put(51,92){ { \textsf{Pair A}}}
    \put(51,18){ { \textsf{Pair B}}}
    \put(209,88){ { \textsf{Pair A}}}
    \put(209,15){ { \textsf{Pair B}}}
   \end{picture}
   \caption{Analysis of the system V736 And. \textit{Left:} Light curves of pairs A and B from the TESS satellite. \textit{Right:} ETV analysis of both pairs, with all the data collected for the star from various surveys. Larger symbols represent data taken with a higher weight because of their better-quality data (e.g. TESS). The filled dots are the primary eclipses; the open circles are for the secondary ones. The solid red curve for pair A stands for the classical light travel time effect fit, while for pair B such a fit with an opposite shape is plotted as a dashed blue line (obviously not fitting the data).}
   \label{Fig_V736And}
  \end{figure}

The star was first mentioned as a doubly eclipsing system showing two sets of eclipses by \cite{2025MNRAS.538.1160K}, who derived its two periods to  be 0.3596225 days and 0.3069135 days. The two sets of eclipsing binaries are obviously contact types, as expected from their short orbital periods. However, pair A is obviously the dominant one, with its rather deep eclipses, while pair B shows only shallow eclipses with depths of only a few hundredth of magnitude. See Fig. \ref{Fig_V736And} for the TESS (Transiting Exoplanet Survey Satellite) light curves of V736~And.

As one can see in Fig. \ref{Fig_V736And}, the long-term evolution of the orbital period is only fairly visible for pair A (red curve), but for pair B its period remains practically constant. Pair A is deep enough that even the data of worse quality from other photometric surveys can provide clear evidence of its ETV variations. The deviations from the linear ephemerides are large for pair A (about 3\% of its orbital period), while nothing similar can be seen for pair B. If one accepts the classical explanation of the 2+2 doubly eclipsing quadruple, then such an ETV amplitude is proportional to the mass ratio of the two pairs. However, according to our modelling of the light curves, pair A is the dominant pair (comparing the luminosity levels of the two binaries), and hence one would expect its ETV amplitude to be smaller. This is just the opposite of what can actually be seen in the ETV curves. I leave the interpretation as an open question for Sect. \ref{explanations}.

\subsection{ASAS J074939-3037.0}

The second system we analysed was ASAS J074939-3037.0 (also known as UCAC4 297-030671 and TIC 150340869). It is also a doubly eclipsing system discovered to show two periods by \cite{2025MNRAS.538.1160K}. Both of its periods (about 0.4417 d and  0.2649 d) are short enough that the system is also composed of two contact binaries. No detailed analysis of this system was found in the literature. There is also a close-by eclipsing binary, Gaia DR3 5598678509597105024, about 5$^{\prime\prime}$ from ASAS J074939-3037.0. However, I believe that it has almost no effect on the quality of the photometry of the main target or its analysis due to the fact that this one additional eclipsing binary is about six magnitudes fainter than our main target.

  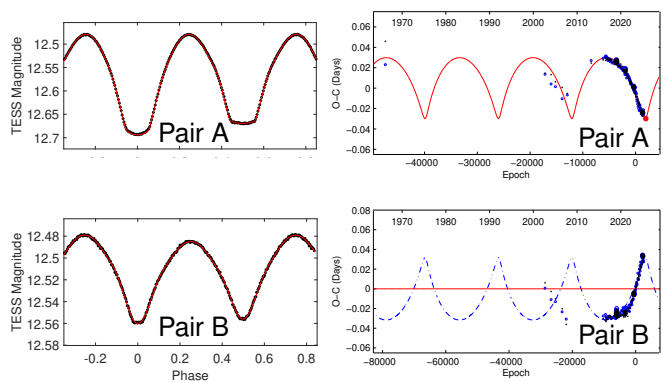
\begin{figure}
  \centering
  \begin{picture}(380,140)
   \put(1,75){\includegraphics[width=0.22\textwidth]{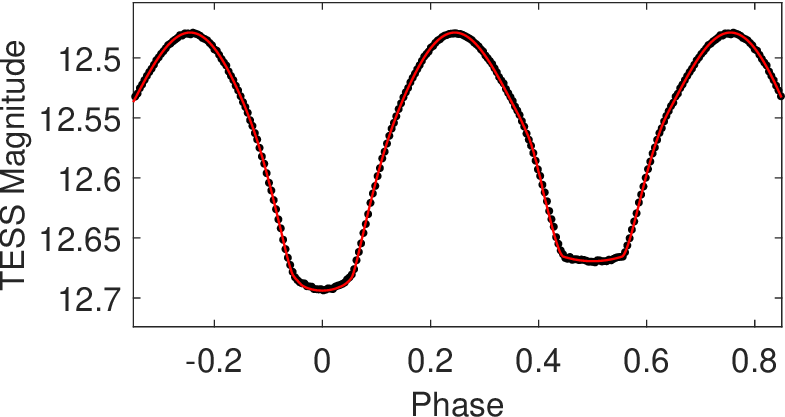}}
   \put(1,1){\includegraphics[width=0.22\textwidth]{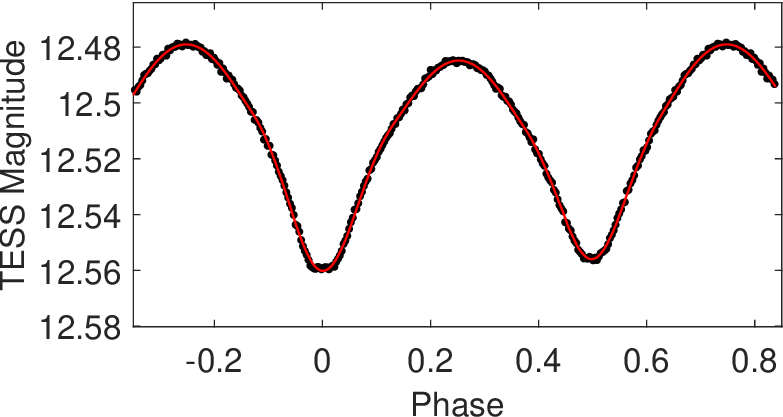}}
   \put(120,0){\includegraphics[width=0.243\textwidth]{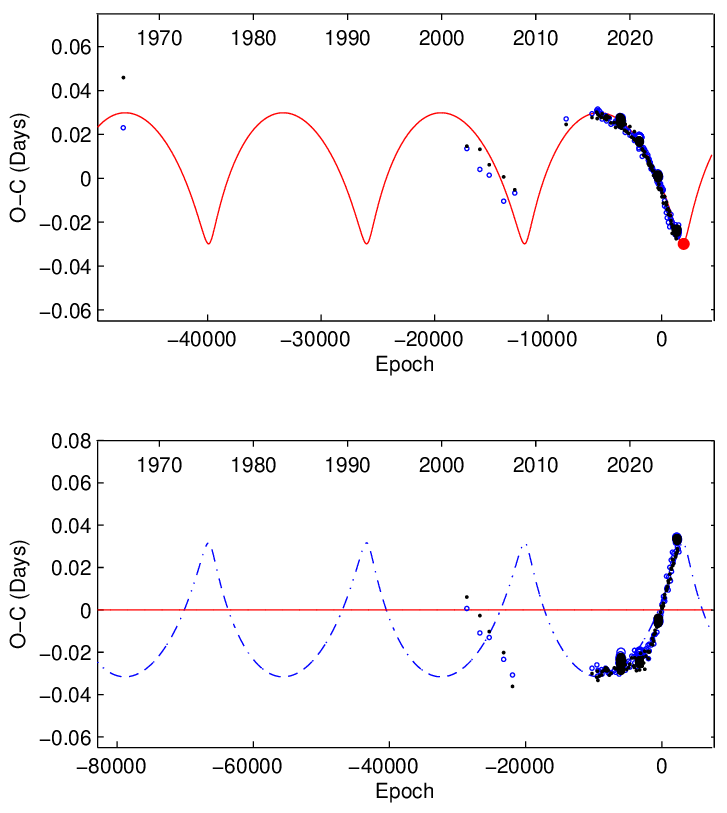}}
    \put(51,91){ { \textsf{Pair A}}}
    \put(51,18){ { \textsf{Pair B}}}
    \put(209,89){ { \textsf{Pair A}}}
    \put(209,15){ { \textsf{Pair B}}}
   \end{picture}
   \caption{Same as Fig. 1 but for the system ASAS J074939-3037.0. The red dot represents the new dedicated observation.}
   \label{Fig_ASAS074939}
  \end{figure}

All available data were collected for the subsequent analysis. The light curve was fitted using TESS photometry, and both pairs were successfully disentangled into separate periodic signals. As one can see from the fits in Fig. \ref{Fig_ASAS074939}, the curves show a contact shape for the A and B pairs. However, pair A is the dominant one, as can be inferred from the luminosity fraction of the third light values given in Table \ref{TabLCfit}. 

The long-term period variations of both pairs were also studied using older data from surveys such as ASASsn, ASAS, and ATLAS, and even digitalised photographic plates via the APPLAUSE project. This was possible mainly due to the fact that pair A shows deep enough eclipses and that the star itself is relatively bright. The ETV can be clearly detected for pair A. However, its shape for pair B remains unexplained. One would expect, based on the classical explanation, that the ETV of pair B would simply be of the opposite shape of A, but this is only partly true and only for the most recent data. The older data clearly show a very different behaviour.

\subsection{OGLE BLG-ECL-123507}

In our sample of stars, the system OGLE BLG-ECL-123507 (also known as OGLE BLG194.6 77405 and TIC 128184860) is of the earliest spectral type, and it is also the faintest. Both of its binaries have orbital periods greater than one day (1.017 d and 2.949 d), and both light curves definitely have a detached-like shape. The system was discovered as a doubly eclipsing system by \cite{2023A&A...674A.170A}.

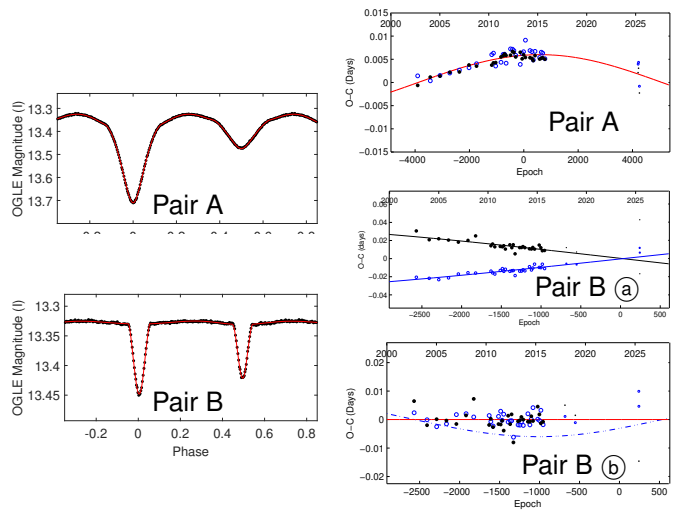
\begin{figure}
  \centering
  \begin{picture}(380,180)
   \put(1,95){\includegraphics[width=0.22\textwidth]{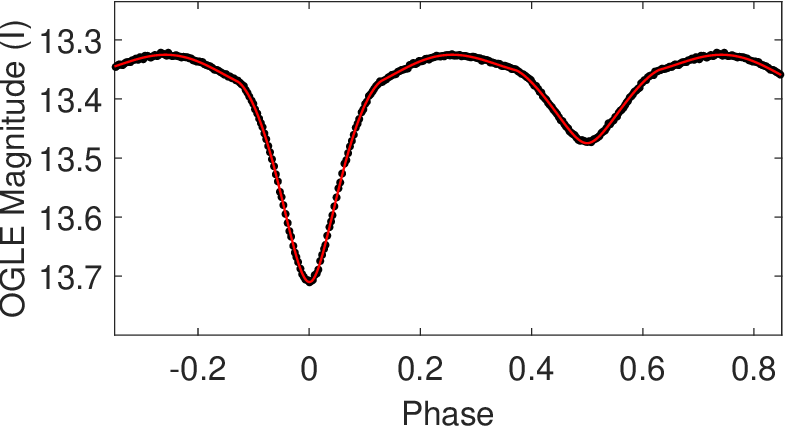}}
   \put(1,21){\includegraphics[width=0.22\textwidth]{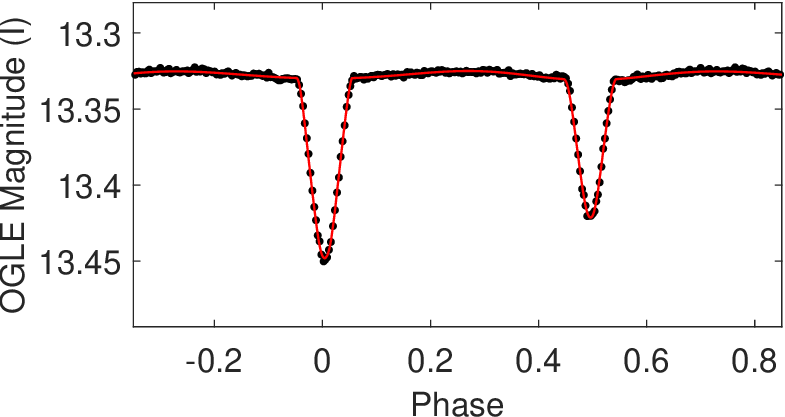}}
   \put(125,125){\includegraphics[width=0.24\textwidth]{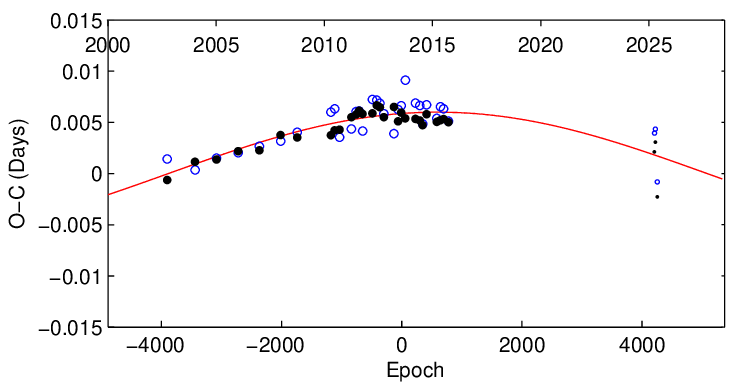}}
   \put(130,69){\includegraphics[width=0.23\textwidth]{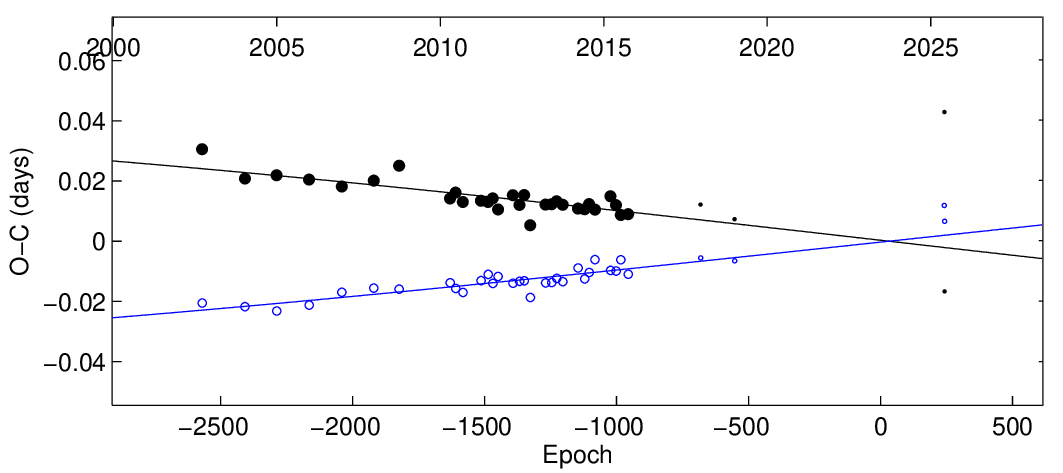}}
   \put(125,0){\includegraphics[width=0.24\textwidth]{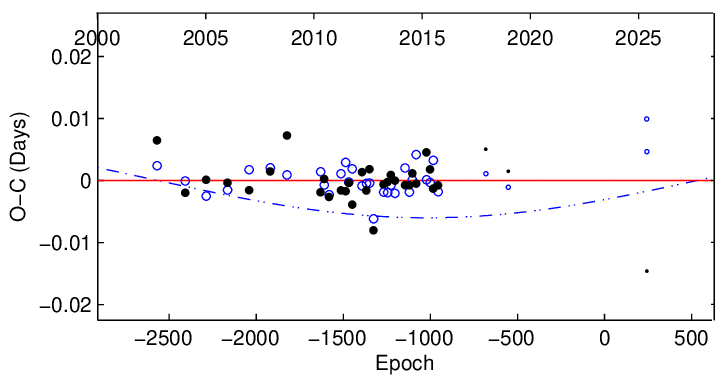}}
    \put(51,112){ {\textsf{Pair A}}}
    \put(51,38){ {\textsf{Pair B}}}
    \put(200,143){ {\textsf{Pair A}}}
    \put(195,81){ {\textsf{Pair B \textcircled{\scriptsize{a}}}}}
    \put(190,14){ {\textsf{Pair B \textcircled{\scriptsize{b}}}}}
   \end{picture}
   \caption{Similar to Fig. 1 but for the system OGLE BLG-ECL-123507. The figures of the A and B pairs are accompanied by the ETV diagram of the slow apsidal motion of pair B (panel \textcircled{\scriptsize{a}}). Panel \textcircled{\scriptsize{b}} was plotted after subtraction of the long-term apsidal evolution.}
   \label{Fig_OGLE}
  \end{figure}

This system is the only one where OGLE data were used instead of TESS data for the modelling of the light curves of both pairs. This is due to the faintness of the star and the significant blending of nearby brighter stars in large TESS pixels causing the TESS light curves to have only very small eclipse amplitudes (about one hundred times lower than from OGLE). The fits of both light curves are given in Fig. \ref{Fig_OGLE}. The resulting values of the light curve parameters are given in Table \ref{TabLCfit}. As one can see, the dominant pair also seems to be pair A.

Concerning the period analysis of both pairs, I followed the same procedure as for the previous systems. Pair A shows quite noticeable period variations, but no such variations are detected for pair B after subtraction of its slow apsidal motion. Both these ETVs are plotted in Fig. \ref{Fig_OGLE}. The eccentricity of pair B is about 0.04, and its apsidal motion has a period of about 200 years. 

\subsection{ASASSN-V J180818.54-684329.4}

The system ASASSN-V J180818.54-684329.4 (also known as TYC 9290-2490-1) is a relatively bright system, but it has not been studied in detail before. As a doubly eclipsing system, the star was first mentioned by \cite{2025MNRAS.538.1160K}, who determined its eclipsing periods to be 0.3392 d and 3.323 d.

\begin{figure}
  \centering
  \begin{picture}(380,140)
   \put(1,75){\includegraphics[width=0.22\textwidth]{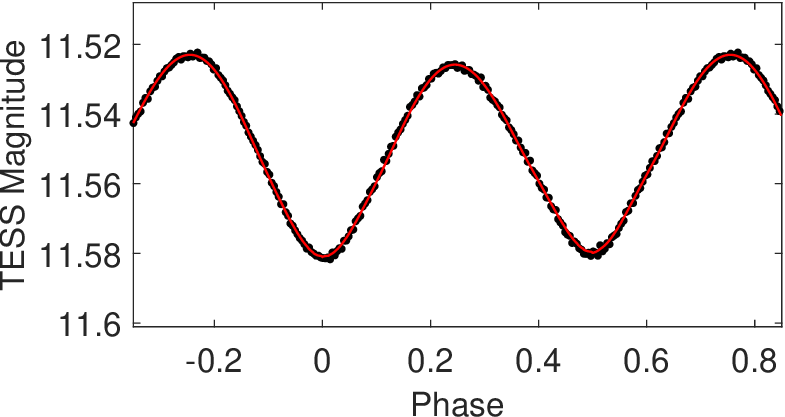}}
   \put(1,1){\includegraphics[width=0.22\textwidth]{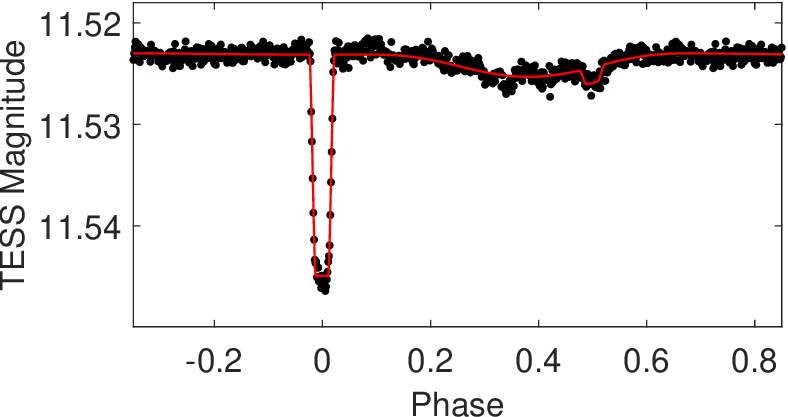}}
   \put(120,0){\includegraphics[width=0.243\textwidth]{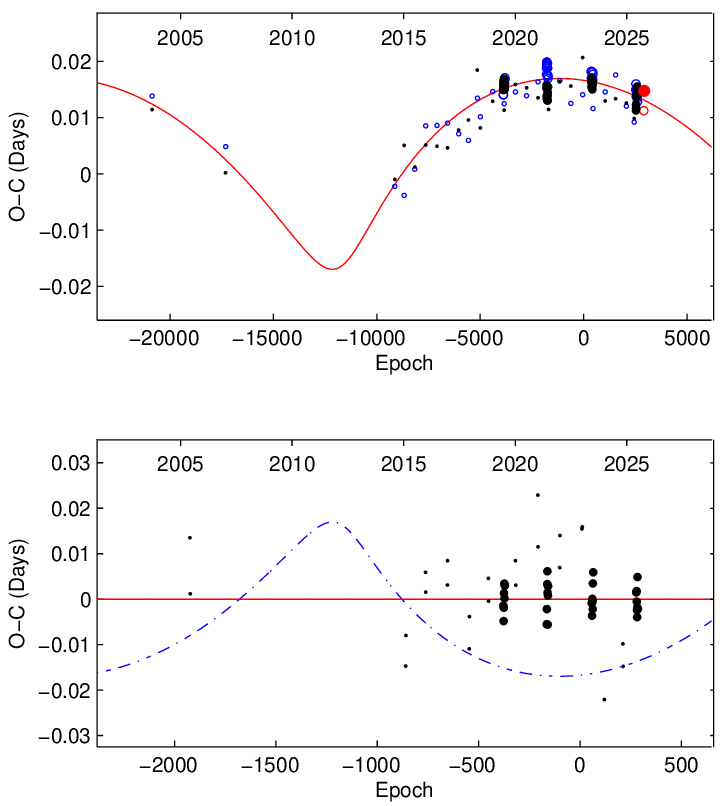}}
    \put(51,91){ { \textsf{Pair A}}}
    \put(51,18){ { \textsf{Pair B}}}
    \put(205,94){ { \textsf{Pair A}}}
    \put(155,18){ { \textsf{Pair B}}}
   \end{picture}
   \caption{Same as Fig. 1 but for the system ASASSN-V J180818.54-684329.4.}
   \label{Fig_ASAS180818}
  \end{figure}
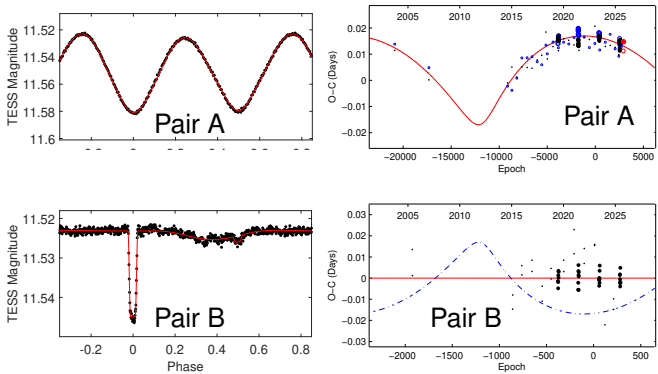

Despite the small amplitude of its photometric variations, both light curves were analysed. The orbital period of pair B is still rather uncertain since the secondary eclipse is very shallow (only about 1~mmag deep). However, I believe that the current period of 3.3~d is the correct one instead of the double value of 6.6~d. The results of our fitting are given in Table \ref{TabLCfit}, while the two fits for pairs A and B are given in Fig. \ref{Fig_ASAS180818}. As one can see, pair A dominates in luminosity. However, additional light not originating from either of these two binaries  is also obviously present in the system (see the large values of the third lights for both pairs). Due to the asymmetries of both light curves, there must also be additional spots on the surfaces of the stars.

The same procedure as for the previous stars was followed for the period analysis. The ETVs are clearly visible for pair A and have an amplitude of more than 20 minutes. However, such a large amplitude cannot also be easily found for pair B, as is clearly shown in Fig. \ref{Fig_ASAS180818}. Pair B more or less follows the same linear ephemeris without any additional curvature.

\subsection{CRTS J191726.4-543540}

The last system we analysed is CRTS J191726.4-543540 (also known as UCAC4 178-222491). The star was never studied in detail, but both of its eclipsing periods were detected by \cite{2025MNRAS.538.1160K}. Quite surprisingly, \textit{Gaia} DR3 lists two sources around the coordinates of CRTS J191726.4-543540 (separated by 0.36$^{\prime\prime}$), and they have about the same brightnesses in the \textit{Gaia} G filter. However, the parallax and proper motion is missing for one star, while the other one shows a large re-normalised unit weight error value of about 7.4. Such a large value should indicate a binary movement, or other peculiarities. Another interesting finding about this system is the fact that the two eclipsing periods are almost exactly in a 1:9 ratio (only about 0.07\% away from the period resonance).

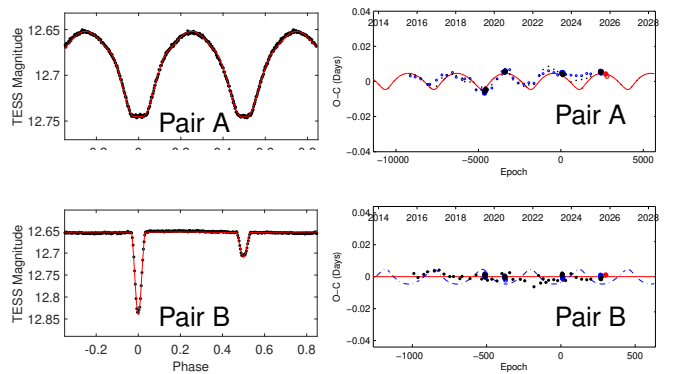
\begin{figure}
  \centering
  \begin{picture}(380,140)
   \put(1,75){\includegraphics[width=0.22\textwidth]{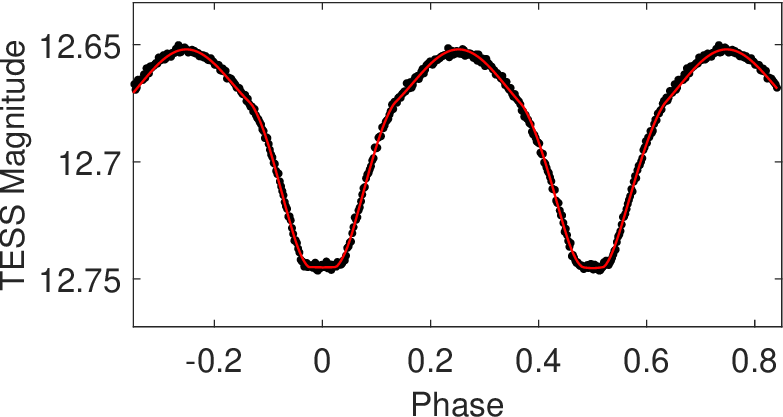}}
   \put(1,1){\includegraphics[width=0.22\textwidth]{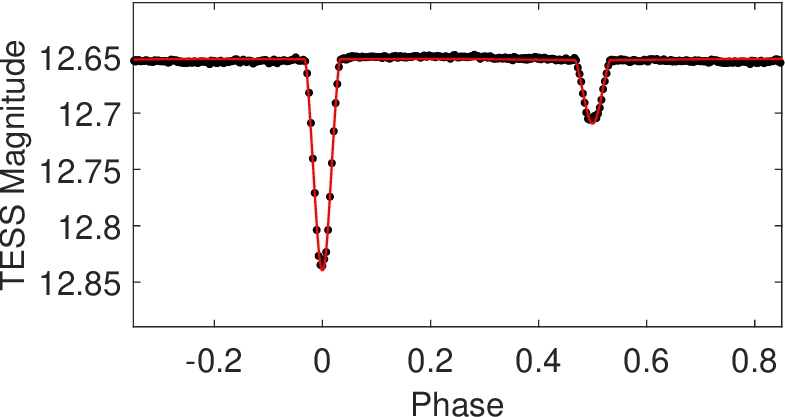}}
   \put(120,0){\includegraphics[width=0.24\textwidth]{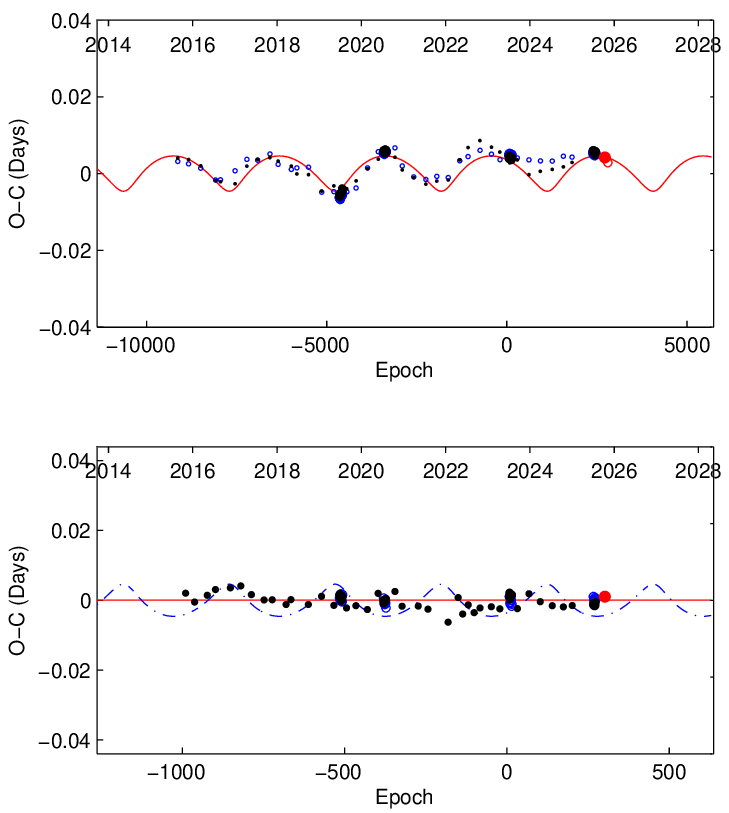}}
    \put(51,91){ { \textsf{Pair A}}}
    \put(51,18){ { \textsf{Pair B}}}
    \put(200,94){ { \textsf{Pair A}}}
    \put(200,18){ { \textsf{Pair B}}}
   \end{picture}
   \caption{Same as Fig. 1 but for the system CRTS J191726.4-543540.}
   \label{Fig_CRTS191726}
  \end{figure}

As for the previous stars, the analysis was carried out on the TESS data for the light curve. Both pairs were analysed, resulting in the light curve parameters given in Table \ref{TabLCfit}, and the final fits are plotted in Fig. \ref{Fig_CRTS191726}. The A and B pairs share a similar brightness. It seems that the fit of pair A is slightly asymmetric, which is even more visible when plotting the light curve of this pair in different sectors of the TESS data. A very small asymmetry is also visible for pair B (especially when comparing the levels of brightness in the two quadratures).

The ETV analysis was done in a similar way as for the previous systems. However, we observed a prominent variation for pair A, at a level of about 7 minutes in amplitude, and period of about 2.5 yr. Nevertheless, this variation is not strictly periodic: it is a quasi-periodic ETV signal. No such variation can be seen for pair B. Its period has remained constant over the past 10 years.

\section{Several hypothetical explanations} \label{explanations}

The individual systems and their long-term period variations are very different from each other, but they may share some similarities. The systems V736 And and ASASSN-V J180818.54-684329.4 hardly show one period of the ETV variation. Hence, their periodicities may also be rather different. For the system OGLE BLG-ECL-123507, one cannot even be sure that the true variation is not a small part of the parabolic trend due to mass transfer (a near semi-detached configuration). For the last system, CRTS J191726.4-543540, one can even see that the individual variations in the ETVs are not strictly periodic, and their shape changes over time. Any such quasi-periodic modulation cannot be explained by simple orbital movement and the light travel time effect. Both pairs of the system ASAS J074939-3037.0 show very complex variations, but on the other hand they are both contact types. Remarkably, for the systems with one obvious contact and one detached system, the ETV is visible for the contact system.

To describe these systems and the strange behaviour of the ETV for both pairs, several possible explanations have been put forwards. These hypothetical interpretations are briefly summarized below. However, this only represents a list of different hypotheses, and it should not be taken as a list organized from the most to the least probable explanations.

\begin{itemize}
    \item One possible explanation for the non-existence of any ETV signal for the secondary pair (pair B) is the mass of pair B being too high compared with the mass of pair A. The individual ETV amplitudes for pairs A and B are proportional to their mass ratio ($M_B/M_A$), at least for the classical geometrical light travel time effect. Therefore, one can also expect that, if pair B has a very high mass, its amplitude would then be so small that it could not be detected even with the quality of today's data. However, I should also mention that such an explanation is highly improbable due to the fact that from the light curve solution, the luminosity fractions of both pairs were also derived, and such a luminosity fraction should be in accordance with the mass ratio of the two pairs. Assuming some non-trivial under-luminous or degenerate objects would only provide a very dubious explanation, so I consider this explanation to be improbable.
    \item  Another explanation would be that a more complex dynamical interaction occurs in these systems. In the case of dynamical interactions, the strict opposite shape of the ETV of pair A versus B disappears. In general, the dynamical effects can sometimes produce very complicated variations. However, there is one aspect that also makes this explanation odd. These effects are only present in systems with very small ratios of their inner and outer periods, i.e. usually stars orbiting on much shorter periods than our systems (see e.g. \citealt{2025ApJ...985..213P}). Here, our detected periods of pair A's variations are of the order of years, and hence its dynamical effects would be negligible. Moreover, the period of pair B remains practically constant over the studied time span.
    \item A third possible explanation is based on the asymmetries in the light curve shapes of the (mostly contact) eclipsing pairs and their evolution in time. Due to these asymmetries, the apparent period variations can occur in these (light-curve distorted) systems. Such a finding would be in accordance with a similar finding recently published by \cite{2025A&A...695A.209B}, who also report complex period variations for several stars of contact configuration. The role of asymmetric light curves and their role in producing ETV signals has been studied elsewhere in the past (see e.g. \citealt{2004MNRAS.351..110W}, \citealt{2011IBVS.5991....1Z}, \citealt{2013ApJ...774...81T}). We note that one of our systems, namely OGLE BLG-ECL-123507, is not a contact type. On the other hand, the asymmetries in the light curve can also be present in this type of system due to the evolving spots or other phenomena. 
    \item Finally, the last possible explanation would be that the system is not a 2+2 quadruple. One possibility is that the system has a (2+1) + 2 architecture. This would mean that the period variation of pair A is due to the movement around a hidden body, while pair B orbits this inner triple on a much longer period that cannot be detected yet. There is still a place for such an orbit in the parameter space in between (these inner and outer orbits) for such a variation (lower limit given by the stability limit, higher limit given by the non-resolution of the A-B pair as a double on the sky). However, any such putative fifth component would be only slightly visible in our analysis (for example, as additional light). On the other hand, such a configuration is not anything exceptional, as demonstrated, for instance, in the MSC (Multiple Star Catalogue) by \cite{2018ApJS..235....6T}. In MSC, quintuple systems of this architecture exist, and the period distribution of the long orbits sometimes reaches up to $10^6$ -- $10^7$ days.
\end{itemize}

What should be mentioned is the fact that the true origin of such variation should in principle be a combination of several of these effects. The explanation may also be different for different stars. 

One possible solution would be to perform long-term spectroscopic monitoring of these systems and to try to detect the changes of the radial velocities of the inner A pair's centre of mass. Such a movement would be in accordance with the ETV of pair A in the case of the last our explanation. Unfortunately, such monitoring would be quite challenging, even with current techniques, due to the low luminosities, short orbital period, and blending of all four components in the spectra.

Having no solid proof in hand, I can only speculate about the true origin of the ETV changes detected. At this point I believe the most probable explanation is a combination of asymmetry-driven ETVs (quasi-periodic) and very long orbits of the A-B pair, which cannot be detected at the present time.

\section{Conclusions}

About one thousand doubly eclipsing candidates, i.e. stars with two different eclipsing periods, have been found to date. Among these candidates, about 60 systems were proved to have two sets of binaries that orbit around each other, i.e. true 2+2 quadruples. 

Until this study, only two groups of doubly eclipsing systems had been found. In one group, the systems show periodic ETVs for both pairs but with opposite shapes, revealing the bound nature of these quadruples. In the second group, there are many other systems showing two sets of eclipses, but visible ETVs cannot be detected. Hence, the systems in this second group can only be considered candidates, as there is no proof of the mutual movement of the pairs in the systems. Their respective periods are probably too long, but another explanation could be a face-on orientation of the mutual orbit.

In the present study I have introduced for the first time systems that defy the previous easy division. The systems show ETVs for only one pair, while the period of the other pair remains constant or shows a different variation. No clear explanation for such a behaviour has been determined, but the most probable is the time-changing asymmetry of the light curve of the contact inner binary. However, only further analysis and the collection of new data will reveal the true nature of these changes.

 \medskip

\begin{acknowledgements}
I do thank the ZTF, ASAS-SN, SWASP, Atlas, OGLE, and TESS teams for making all of the observations easily public available. I am also grateful to the ESO team at the La Silla Observatory for their help in maintaining and operating the Danish 1.54m telescope. The research of P.Z. was also supported by the project {\sc Cooperatio - Physics} of Charles University in Prague. This research made use of Lightkurve, a Python package for TESS data analysis \citep{2018ascl.soft12013L}. Funding for APPLAUSE has been provided by DFG (German Research Foundation, Grant), Leibniz Institute for Astrophysics Potsdam (AIP), Dr. Remeis Sternwarte Bamberg (University Nuernberg/Erlangen), the Hamburger Sternwarte (University of Hamburg) and Tartu Observatory. Plate material also has been made available from Th\"uringer Landessternwarte Tautenburg, and from the archives of the Vatican Observatory.  This research has made use of the SIMBAD and VIZIER databases, operated at CDS, Strasbourg, France and of NASA Astrophysics Data System Bibliographic Services. This work has made use of data from the European Space Agency (ESA) mission {\it Gaia} (\url{https://www.cosmos.esa.int/gaia}), processed by the {\it Gaia} Data Processing and Analysis Consortium (DPAC, \url{https://www.cosmos.esa.int/web/gaia/dpac/consortium}). Funding for the DPAC has been provided by national institutions, in particular the institutions participating in the {\it Gaia} Multilateral Agreement.
\end{acknowledgements}

\begin{appendix}

\section{Appendix table}

\begin{table*} 
\caption{Derived parameters for the two inner binaries, A and B.}
 \label{TabLCfit}
  \centering \scalebox{0.75}{
\begin{tabular}{c | c | c | c | c | c }
   \hline\hline
 \multicolumn{1}{c|}{System } &        V736 And          &   ASAS J074939-3037.0    &   OGLE BLG-ECL-123507   & ASASSN-V J180818.54-684329.4 &   CRTS J191726.4-543540   \\ \hline
  \multicolumn{1}{c|}{ }      &   \multicolumn{5}{c}{{\sc p a i r \,\, A}} \\ \hline
  $JD_0$ [2450000+\dots]      &  8508.1999 $\pm$ 0.0006  & 9998.2019 $\pm$ 0.0011   & 6500.9120 $\pm$ 0.0007  &   9973.0432 $\pm$ 0.0014     & 10108.7881 $\pm$ 0.0005   \\
  $P$ [d]                     & 0.3596183 $\pm$ 0.0000002& 0.4417150 $\pm$ 0.0000003&1.0169733 $\pm$ 0.0000004&  0.3391949 $\pm$ 0.0000003   & 0.3128589 $\pm$ 0.0000002 \\
  $i$ [deg]                   &    76.76 $\pm$ 0.26      &    88.99 $\pm$ 0.18      &     77.94 $\pm$ 0.14    &      49.17 $\pm$ 0.47        &   72.47 $\pm$ 0.21       \\
 $q=\frac{M_2}{M_1}$          &    1.17 $\pm$ 0.05       &     0.14 $\pm$ 0.02      &     0.83 $\pm$ 0.02     &      0.98 $\pm$ 0.05         &    0.12 $\pm$ 0.02       \\
 $T_1$ [K]                    &     5805 (fixed)         &     6281 (fixed)         &     7970 (fixed)        &      5911 (fixed)            &     5821 (fixed)         \\
 $T_2$ [K]                    &     6131 $\pm$ 47        &     6229 $\pm$ 34        &     4783 $\pm$ 38       &      5828 $\pm$ 69           &     5170 $\pm$ 43        \\
 $R_1/a$                      &    0.469 $\pm$ 0.006     &    0.570 $\pm$ 0.004     &    0.379 $\pm$ 0.003    &     0.381 $\pm$ 0.002        &   0.538 $\pm$ 0.004      \\
 $R_2/a$                      &    0.493 $\pm$ 0.005     &    0.247 $\pm$ 0.003     &    0.344 $\pm$ 0.005    &     0.379 $\pm$ 0.002        &   0.202 $\pm$ 0.005      \\
 $L_1$ [\%]                   &    25.5 $\pm$ 0.7        &     57.7 $\pm$ 0.5       &     46.3 $\pm$ 0.7      &     19.0 $\pm$ 0.5           &    48.3 $\pm$ 0.8        \\
 $L_2$ [\%]                   &    33.8 $\pm$ 0.9        &      9.1 $\pm$ 0.4       &     11.4 $\pm$ 0.5      &     21.4 $\pm$ 0.5           &     4.1 $\pm$ 0.2        \\
 $L_3$ [\%]                   &    40.7 $\pm$ 1.1        &     33.2 $\pm$ 0.8       &     42.3 $\pm$ 1.0      &     59.6 $\pm$ 2.2           &    47.6 $\pm$ 1.7        \\ \hline
  \multicolumn{1}{c|}{ }      & \multicolumn{5}{c}{{\sc p a i r \,\, B}}\\ \hline
 $JD_0$ [2450000+\dots]       &  8508.3357 $\pm$ 0.0008  &  9999.1568 $\pm$ 0.0012  &  9963.968 $\pm$ 0.0026  &  9911.5134 $\pm$ 0.0020      & 10109.5188 $\pm$ 0.0017  \\
 $P$ [d]                      & 0.3069129 $\pm$ 0.0000004& 0.2648485 $\pm$ 0.0000004&2.9487665 $\pm$ 0.0000031& 3.3233982 $\pm$ 0.0000052    & 2.8178296 $\pm$ 0.0000020 \\
 $i$ [deg]                    &     53.98 $\pm$ 0.51     &     64.15 $\pm$ 0.62     &     80.78 $\pm$ 0.12   &      87.51 $\pm$ 0.40         &   83.55 $\pm$ 0.47      \\
 $q=\frac{M_2}{M_1}$          &    0.97 $\pm$ 0.04       &    1.01 $\pm$ 0.03       &      1.00 (fixed)      &       1.00 (fixed)            &    1.00 (fixed)         \\
 $T_1$ [K]                    &     5805 (fixed)         &     6281 (fixed)         &      7970 (fixed)      &       5911 (fixed)            &    5821 (fixed)         \\
 $T_2$ [K]                    &     5507 $\pm$ 48        &     6242 $\pm$ 67        &      7669 $\pm$ 53     &       3804 $\pm$ 81           &    4696 $\pm$ 39        \\
 $R_1/a$                      &    0.397 $\pm$ 0.004     &    0.380 $\pm$ 0.003     &    0.171 $\pm$ 0.002   &      0.035 $\pm$ 0.002        &  0.119 $\pm$ 0.002      \\
 $R_2/a$                      &    0.392 $\pm$ 0.003     &    0.383 $\pm$ 0.003     &    0.172 $\pm$ 0.002   &      0.119 $\pm$ 0.004        &  0.116 $\pm$ 0.002      \\
 $L_1$ [\%]                   &    16.0 $\pm$ 0.4        &    14.3 $\pm$ 0.3        &      25.3 $\pm$ 0.3    &       2.0 $\pm$ 1.1           &   38.5 $\pm$ 0.8        \\
 $L_2$ [\%]                   &    11.5 $\pm$ 0.4        &    12.8 $\pm$ 0.7        &      17.4 $\pm$ 0.3    &       1.5 $\pm$ 0.8           &   11.4 $\pm$ 0.6        \\
 $L_3$ [\%]                   &    72.5 $\pm$ 0.9        &    72.9 $\pm$ 1.4        &      57.3 $\pm$ 0.8    &      96.5 $\pm$ 3.4           &   50.1 $\pm$ 2.0        \\
 \noalign{\smallskip}\hline\hline
\end{tabular}} \\
\end{table*}

\end{appendix}

\end{document}